# Increase of magnetic hyperthermia efficiency due to dipolar interactions in low anisotropy magnetic nanoparticles : theoretical and experimental results.

B. Mehdaoui, R. P. Tan, A. Meffre, J. Carrey\*, S. Lachaize, B. Chaudret and M. Respaud

Université de Toulouse; INSA; UPS; LPCNO (Laboratoire de Physique et Chimie des Nano-Objets), 135 avenue de Rangueil, F-31077 Toulouse, France and
CNRS; UMR 5215; LPCNO, F-31077 Toulouse, France

**Abstract:**

When magnetic nanoparticles (MNPs) are single-domain and magnetically independent, their magnetic properties and the conditions to optimize their efficiency in magnetic hyperthermia applications are now well-understood. However, the influence of magnetic interactions on magnetic hyperthermia properties is still unclear. Here, we report hyperthermia and high-frequency hysteresis loop measurements on a model system consisting of MNPs with the same size but a varying anisotropy, which is an interesting way to tune the relative strength of magnetic interactions. A clear correlation between the MNP anisotropy and the squareness of their hysteresis loop in colloidal solution is observed : the larger the anisotropy, the smaller the squareness. Since low anisotropy MNPs display a squareness higher than the one of magnetically independent nanoparticles, magnetic interactions enhance their heating power in this case. Hysteresis loop calculations of independent and coupled MNPs are compared to experimental results. It is shown that the observed features are a natural consequence of the formation of chains and columns of MNPs during hyperthermia experiments: in these structures, when the MNP magnetocristalline anisotropy is small enough to be dominated by magnetic interactions, the hysteresis loop shape tends to be rectangular, which enhance their efficiency. On the contrary, when MNPs do not form chains and columns, magnetic interactions reduces the hysteresis loop squareness and the efficiency of MNPs compared to independent ones. Our finding can thus explain contradictory results in the literature on the influence of magnetic interactions on magnetic hyperthermia. It also provides an alternative explanation to some experiments where an enhanced specific absorption rate for MNPs in liquids has been found compared to the one of MNPs in gels, usually interpreted with some contribution of the brownian motion. The present work should improve the understanding and interpretation of magnetic hyperthermia experiments.

**Main Text:**

1. Introduction

Maximizing the specific absorption rate (SAR) of nanoparticles in magnetic hyperthermia and interpreting experimental results is a complex task since the magnetic properties of assemblies of magnetic nanoparticles (MNPs) depend of a large number of parameters. When MNPs can be considered as single-domain and magnetically independent, underlying mechanisms are now well-understood [1, 2]. However, several experimental and theoretical studies have shown that the presence of magnetic interactions between MNPs dramatically influence their SAR [3, 4, 5, 6, 7, 8, 9, 10]. When MNPs are small enough to clearly have a superparamagnetic behaviour, the influence of magnetic interactions is relatively simple: magnetic interactions increase their effective anisotropy [3]; in most cases, this is expected to enhance their susceptibility and SAR by bringing them closer to the superparamagnetic/ferromagnetic transition. However, such NPs are not the most suitable for magnetic hyperthermia and larger NPs close to the superparamagnetic-ferromagnetic transition or well into the ferromagnetic regime have been shown to be more efficient [1]. Influence of magnetic interactions in these latter is more complex and apparently contradictory results have been found. Experimentally, an increase [5], a decrease [7, 8] or a non-monotonic [10] variation of SAR with interactions have been reported. In our group, quantitative analysis of SAR measurements on two different types of single-domain MNPs (FeCo and Fe) had led us to conclude that magnetic interactions decrease SAR [11, 12]. From a theoretical point of view, most theoretical works agree that the general trend is that SAR decreases with interactions [3, 8, 9, 10] although a limited increase in a restricted range of NP concentration has also been reported [9, 10]. Except Ref. [9], theoretical studies have been performed on isotropic assemblies of NPs, despite that it is well-known that applying a magnetic field to a ferrofluid induces the formation of chains or columns [13, 14].

The present work aims at presenting a set of both theoretical and experimental results related to the influence of magnetic interactions in magnetic hyperthermia efficiency. The measurements were performed on an ensemble of samples in which the size of the MNPs was mainly kept constant but their anisotropy could be varied by changing their composition (Fe, FeCo, $Fe_xC_y$) and their intimate structure. This is a unique way to tune the relative strength of magnetic interactions and study their influence on hyperthermia properties. The strength of this work lies in the following features: i) high frequency hysteresis loops were measured in the same conditions as hyperthermia. The hysteresis loops contain more information on the sample magnetic properties than basic SAR measurements and are thus a precious tool to understand the physics of magnetic hyperthermia. ii) Experimental results were compared with hysteresis loop calculations of independent and coupled MNPs. In particular, the fact that MNPs form chains or columns has been taken into account and is shown to be a key element to understand experimental results. iii) These MNPs display optimized characteristics for magnetic hyperthermia and very large SARs, so these results are of real interest for eventual applications.

Our different approaches converge to the same conclusion that increasing the strength of magnetic interactions can improve the efficiency of MNPs in colloidal solution, *i.e.* when the nanoparticles are free to arrange in chains and columns. However, this improvement only occurs when dipolar interactions are sufficiently strong to overcome the magnetocrystalline anisotropy of NPs and increase the squareness of their hysteresis loop. This is precisely the case for low anisotropy NPs, which are the preferred candidates for magnetic hyperthermia. However, our numerical simulations evidence that the conclusion is opposite if the nanoparticles are in isotropic conditions, *i.e.* if they are not able to move because they are blocked in a gel or in a cell for instance. Thus our results are not contradictory with previous theoretical works but only more general; they provide an alternative explanation for the increase of SAR of nanoparticles in liquid compared to a gel, which has been often attributed in the literature to the contribution of Brownian motion. Finally, they should improve the understanding of both *in vivo* and *in vitro* hyperthermia experiments and of the mechanisms of magnetic hyperthermia.

## 2. Methods

Samples under study are colloidal solutions composed of Fe, $Fe_xC_y$, $Fe@Fe_xC_y$ (an iron core surrounded by a $Fe_xC_y$ shell) and FeCo nanoparticles. Their synthesis methods and detailed structural characteristizations have been described in Refs. [11] [15] and [16]. Mean diameter and size distribution of the different samples were measured by transmission electron microscopy (TEM). Standard magnetic characterizations were performed using a SQUID magnetometer on a powder. For SAR and high-frequency hysteresis loop measurements, 10 mg of powder were diluted in 0.5 ml of mesitylene inside a schlenck tube filled with Ar to protect the samples against oxidation. SAR and high-frequency hysteresis loop measurements were performed on the same sample. SAR measurements were performed on a home-made electromagnet specially designed for hyperthermia experiments [17]. High-frequency hysteresis loops were performed using a setup described elsewhere [18]. Hysteresis loops measured on this setup have previously shown to be consistent with SAR measurements [18]. Hysteresis loop calculations of magnetically independent MNPs have been described in details in Ref. [1] and fully tested; they are able to calculate the hysteresis loop area $A$ for an assembly of magnetically independent spherical uniaxial NPs with their anisotropy axis randomly distributed in space or aligned with the magnetic field. Hysteresis loop calculations of coupled nanoparticles are based on the solving of the Landau-Lifshitz-Gilbert (LLG) equation. Details on the model used and results on the influence of magnetic interactions on the magnetic and magnetotransport properties of 2D isotropic assemblies of MNPs can be found in Ref. [19].

## 3. Results

Sample names and characterization results are summarized in Table 1. TEM micrographs are shown in Fig. 1. All the samples are composed of MNPs with a similar diameter, which varies slightly around 13.5 nm. The saturation magnetization per unit mass $\sigma_S$ of sample 1 is 140±8 $Am^2.kg^{-1}$, well below the bulk value (240 $Am^2kg^{-1}$), due to an imperfect alloying of Fe and Co atoms and an amorphous structure of the NPs [20, 21]. Their amorphous structure also explains their very low anisotropy value, previously deduced from hyperthermia experiments [11]. Extensive structural characterization by X-Ray diffraction, high-resolution TEM and Mössbauer spectroscopy on Fe, $Fe_xC_y$, and $Fe@Fe_xC_y$ samples (samples 2-6) have been published in Ref [16]. In this series, changes in the synthesis parameters leads to change in the carbon content and phases of the MNPs. Sample 2 contains pure iron monocrystalline NPs. Sample 3 contains NPs composed of a crystalline $Fe_{2.2}C$ core and a thin amorphous $Fe_{2.5}C$ shell. In samples 4-6, the carbon content and phases vary. Since magnetocristalline anisotropy is very sensitive to exact composition, its value is strongly modulated in the series of Fe, $Fe_xC_y$, and $Fe@Fe_xC_y$ samples. Magnetization saturation is close to the one of bulk Fe for sample 2 (Fe) and samples 4-6 ($Fe@Fe_xC_y$) and is reduced in sample 3 ($Fe_xC_y$).

SAR and high-frequency hysteresis loop measurements have been performed at the same frequency $f$ = 54 kHz. In SAR measurements, the maximum applied magnetic field $\mu_0H_{max}$ was varied between 0 and 60 mT. When a large magnetic field is applied during hyperthermia experiments, it is seen with the naked eye that MNPs self organize into needles, similarly to what was reported in Ref [13]. To evidence it, we have deposited a drop of colloidal solution of Fe MNPs on a TEM grid and let it dry under the application of a 40 mT alternating magnetic field at 54 kHz (see Fig. 2). Although the result obtained cannot be fully representative of what happens during hyperthermia experiments, it still gives an idea of the magnetic field influence on the MNP organization during hyperthermia. In Fig. 3(a), SAR values as a function of the magnetic field measured on the different samples are shown. High-frequency hysteresis loops were measured at $\mu_0 H_{max}$= 42 mT. Raw and normalized hysteresis loops are displayed in Figs. 3(b) and 3(c), respectively. From the area $A$ of these hysteresis loops, it is also possible to calculate a SAR value at 42 mT, using the equation $SAR = Af$. The calculated values are shown in Table 1 along with SAR values deduced from temperature measurements. A reasonable agreement is found between the two methods, similarly to what was found in Ref. [18].

The shape of the hysteresis loops combined with the SAR measurements for each sample lead to very useful information on the magnetic properties of each sample. Except sample 3, all the samples displays a behavior typical of the ferromagnetic regime [1, 11, 12], as evidenced by i) the measurement of widely opened saturated hysteresis loops and ii) an abrupt increase in SAR($\mu_0 H_{max}$) curves followed by a saturation. For sample 3, which is mainly composed of crystalline $Fe_{2.2}C$, the different behavior is probably due to an anisotropy value much higher than the other samples – and then a coercive field higher than our maximum applied field. This is what will be assumed in the remaining of this article.

From the hysteresis loops shown in Figs. 3(a) and 3(b), a second observation can be made : samples 1, 4 and 5 displays a hysteresis loop typical of MNPs with an anisotropy axis oriented with the magnetic field, characterized by both large squareness and remnant magnetization $M_R$. On the contrary, sample 2 displays a shape typical of MNPs with randomly oriented anisotropy axis, characterized by lower squareness and remanence $M_R$. Sample 6 is intermediate between these two behaviors. Sample 3, as discussed above, is not saturated. In the following, we will perform a detailed analysis and interpretation of these various hysteresis loops.

The first step of this analysis is to try to provide an estimation of the effective anisotropy $K_{eff}$ of the MNPs, which will be useful for the rest of the analysis. $K_{eff}$ is derived by two different ways, one based on the hyperthermia experiments and the other one on the hysteresis loops. For the first one, we use the highest slope of the SAR($\mu_0 H_{max}$) function to estimate the anisotropy, similarly to Ref. [12]. For sample 2 and 6 the equation valid for randomly oriented MNPs was used:

$$\mu_0 H_{CHyp} = 0.926 \frac{\mu_0 K_{eff}}{M_S}\left(1-\left[\frac{k_B T}{K_{eff} V}\ln\left(\frac{k_B T}{4\mu_0 H_{CHyp} M_S V f \tau_0}\right)\right]^{0.8}\right), \qquad (1)$$

where $\mu_0 H_{CHyp}$ is the point of highest slope in SAR($\mu_0 H_{max}$) functions, $V$ the MNP volume, $\tau_0$ the frequency factor of the Néel-Brown relaxation time, $M_S$ the magnetization per unit of volume and $T$ the temperature. The experimental $M_S$ value was used. For samples 1, 4 and 5, the following equation was used, which was derived similarly to Equ. (1) using numerical simulations [12] and is valid for MNPs aligned with the magnetic field. Appendix A describes the origin of this equation:

$$\mu_0 H_{CHyp} = 1.85 \frac{\mu_0 K_{eff}}{M_S} \left( 1 - \left[ \frac{k_B T}{K_{eff} V} \ln\left( \frac{k_B T}{4\mu_0 H_{CHyp} M_S V f \tau_0} \right) \right]^{0.5} \right). \quad (2)$$

The second method to deduce $K_{eff}$ uses the coercive field values from the hysteresis loops $\mu_0 H_C$. The following equation was used for samples 2 and 6 [1]:

$$\mu_0 H_C = 0.96 \frac{\mu_0 K_{eff}}{M_S} \left( 1 - \left[ \frac{k_B T}{K_{eff} V} \ln\left( \frac{k_B T}{4\mu_0 H_{max} M_S V f \tau_0} \right) \right]^{0.8} \right) \quad (3)$$

This one was used for samples 1, 4 and 5:

$$\mu_0 H_C = 2 \frac{\mu_0 K_{eff}}{M_S} \left( 1 - \left[ \frac{k_B T}{K_{eff} V} \ln\left( \frac{k_B T}{4\mu_0 H_{max} M_S V f \tau_0} \right) \right]^{0.5} \right) \quad (4)$$

For sample 3, since both $\mu_0 H_C$ and $\mu_0 H_{CHyp}$ are above our maximum available magnetic field, a lower limit for $K_{eff}$ was provided using Equs. (1) and (3). The values found for $K_{eff}$ using these two methods are summarized in Table 1. Please note that these $K_{eff}$ value should not be taken as the one of individual MNPs, but the one of MNPs inside the assembly. In other words, $K_{eff}$ is the value that individual MNPs would have to display a coercive field similar to the one measured. A similar approach was for instance used in Ref. [3] to analyse the influence of magnetic interactions on anisotropy fields and barriers. Except for sample 2, the methods based on hyperthermia measurements and the one based on hysteresis loops converge to $K_{eff}$ values close to each other. For sample 2, the discrepancy probably comes from the large error bar on the $\mu_0 H_{CHyp}$ value due to the smooth shape of the SAR($\mu_0 H_{max}$) function.

We come now to the description of the most interesting features of our set of experimental data. In Figs. 4(a) and 4(b), the values of remnant magnetization $M_R$ normalized by the magnetization at 42 mT $M_{sat}$ and of the slope of the hysteresis loop at the coercive field are plotted as a function of $K_{eff}$. For $K_{eff}$, the value found using magnetic measurements is taken since the error on $\mu_0 H_C$ is lower than on $\mu_0 H_{CHyp}$. It is clear from these figures that a decrease of the MNP anisotropy is correlated with an increase of both normalized remnant magnetization and slope at the coercive field. Another way to characterize this fact is to calculate the squareness of the hysteresis loop, which is directly related to the efficiency of MNPs for their application in hyperthermia [1]. Here, we define the squareness $S$ as

$$S = \frac{A}{4\mu_0 H_{sat} M_{sat}}, \quad (5)$$

where $M_{sat}$ is the magnetization of the sample when it is saturated by a magnetic field $\mu_0 H_{sat}$. For instance, for sample 1, $M_{sat}$ = 33 Am$^2$kg$^{-1}$ and $\mu_0 H_{sat}$ = 12 mT [see Fig. 3(b)]. With this definition, $S$ = 1 for a perfectly square hysteresis loop. In Fig. 4(c), the evolution of the squareness with $K_{eff}$ is plotted for the different samples : an increase of squareness with a decreasing anisotropy is evidenced.

The previous findings can in no case be explained in the framework of magnetically independent single-domain MNPs. To illustrate it, we have performed numerical simulations of hysteresis loops using the model described in Ref. [1], with $\tau_0 = 5 \times 10^{-11}$ s, $M_S = 2 \times 10^6$ A m$^{-1}$, $T$ = 300 K, $f$ = 54 kHz and a varying anisotropy $K_{ind}$. Illustration of the hysteresis loops can be found elsewhere (see Figs. 3(a) and 3(b) in Ref [1]). In Fig. 5, the evolution of the slope at the coercive field, of $M_R/M_S$ and of the squareness are plotted as a function of $K_{ind}$ in the case of MNPs with anisotropy axes oriented with the magnetic field and in the case of anisotropy axes randomly oriented in space. Although the slope at the coercive field varies with anisotropy - similarly to our experimental results- this is not the case for the $M_R/M_S$ ratio and for the squareness. These latter are rather independent of $K_{ind}$ in a large range and drop sharply only when the low-anisotropy nanoparticles are in the superparamagnetic regime. The experimental results are opposite to this tendency since low anisotropy nanoparticles show a strong increase of their $M_R/M_{sat}$ ratio and squareness.

On one other side, experimental results are very well explained by the presence of magnetic interactions. To illustrate it, we have performed simulations based on the solving of the LLG equation at $T$ = 0 K. Single-domain NPs are placed on a square lattice and carry macrospins considered as magnetic point dipoles. More details on the resolution can be found in Ref [19]. In the present study, the following parameters have been considered: $M_S = 2 \times 10^6$ A m$^{-1}$, $D$ = 13 nm, the center-to-center interparticle distance was 15 nm, and the individual anisotropy $K_{ind}$ of the nanoparticles was varied. Hysteresis loops of 3D isotropic assemblies of MNPs as well as anisotropic ones have been studied, with arrays ranging from 6x6x6 to 6x6x60. In Fig. 6, the evolution of the slope at the coercive field, of $M_R/M_S$ and of the squareness are plotted as a function of the nanoparticle individual anisotropy $K_{ind}$. Displaying these data as a function of $K_{eff}$ (i.e. the anisotropy that one would have deduced from the coercive field of the hysteresis loop) does not change the trend of these curves and the conclusion of this paragraph. In Fig. 7 examples of calculated hysteresis loops are shown. In Fig. 8, the magnetic configuration of the MNPs in the remanent state in four typical cases are shown (large/low anisotropy, and isotropic/anisotropic array).

As expected, for large anisotropy MNPs, the magnetic behaviour is independent of the array configuration: hysteresis loops display the typical features of non-coupled assemblies of MNPs at $T$ = 0 K [1] : the $M_R/M_S$ ratio is 0.5 [see Figs 6(a), 7(a) and 7(b)], squareness equals 0.25 [see Figs 6(c)], the coercive field is approximately half of the anisotropy field $\mu_0 H_K = \dfrac{2K_{ind}}{M_S}$ [see Figs. 7(e) and 7(f)] and the magnetic moment of the MNPs are randomly oriented in space in the remanent state [see Figs. 8(b) and 8(d)]. All these features indicate that any influence of dipolar interactions is dominated by the strong anisotropy of individual MNPs.

On the opposite, for low anisotropy MNPs, magnetic properties are extremely sensitive to the configuration of the array. In isotropic arrays, the squareness [see Fig. 6(c)] as well as the $M_R/M_S$ ratio [see Figs. 6(a), 7(c) and 7(e)] both decrease. This is a logical consequence of the fact that when a null magnetic field is applied, a demagnetized state of the assembly is favoured, leading to a wasp-waisted hysteresis loop [see Fig. 7(c)]. Elongated arrays display a very different behaviour : decreasing the anisotropy leads to an increase of the $M_R/M_S$ ratio [see Figs. 6(a), 7(b) and 7(d)] and of the squareness [see Fig. 6(c)]. Fig. 8(c) evidences visually that this originates from the stabilization of the MNP magnetization along the longer axis of the assembly by dipolar interactions (in other words, there is a global easy axis in the assembly resulting from its shape anisotropy). When comparing the various graphs of Fig. 6 with the experimental data shown in Fig. 4, it is clear that the properties observed in hyperthermia experiments are typical of the last case described. We conclude that the formation of chains and/or columns of MNPs during hyperthermia experiments leads to an increase of the squareness of the hysteresis loops – and so to their efficiency - when their anisotropy is small enough.

4. Discussion

In the remaining of this article, we discuss in details our results and make comparisons with the literature. The agreement between our simulations and our experiments is qualitative and not quantitative. Indeed, the anisotropy value for which there is a transition between the coupled and the non-coupled regime is much higher in simulations than experimentally, indicating a weaker influence of magnetic interactions in experiments. This could be due to the fact that our simulations take into account neither the temperature nor the presence of disorder/voids in the assembly and are performed on assemblies of reduced size. A quantitative analysis would require more realistic simulations as well as a much detailed characterization of the chains and columns formed during the experiments.

There is a tricky and unsolved issue emerging from our experimental results. The hysteresis loops measured on all the samples of the present studies display very large squareness, with values up to 0.74 (see Table 1). With such hysteresis loop shape, it should be expected that the measured SAR are much higher than the one measured, getting close to the maximum possible SAR [1]:

$$SAR_{max} \approx 4\mu_0 M_S H_{max} f. \qquad (6)$$

In several previous articles we had attributed reduced SAR compared to what could be expected theoretically to the presence of magnetic interactions, which would have reduced the squareness of hysteresis loops [11, 12]. The present study indicates that this hypothesis was probably wrong since, in our systems, the squareness is on the contrary enhanced by the presence of interactions. This point is clearly demonstrated by the shape of the hysteresis loops. A careful look to Fig. 3(b) and Table 1 indicates a surprising second explanation to the reduction of SAR: the saturation magnetization $M_{sat}$ measured from the high frequency hysteresis loops is for every sample much lower than the $M_S$ measured in SQUID, despite that the hysteresis loop shape seems to indicate that the MNPs are saturated or nearly saturated. One could think that the $M_{sat}$ value provided by our setup is not correct. We cannot completely exclude it; however, the fact that temperature measurements *and* high-frequency hysteresis loops are coherent and give approximately the same

SAR value does not support this hypothesis. It is as if only a part (in the range 20-50%) of the MNPs contributed to the magnetic response and then the heating of the sample. In the case of sample 1, we attributed this discrepancy to the presence of many superparamagnetic MNPs which would not be saturated by the magnetic field [22]. However, in the other samples under study here, there is no trace of such small MNPs. It could also be possible that a part of the MNPs adopt a configuration with their easy axis perpendicular to the magnetic field, where they would not contribute much to the magnetic signal and to the SAR. It has been predicted theoretically that such a configuration occurs when magnetic fields much below the saturating field of the MNPs are applied [23]. Finally, it is also possible that, in the samples, there still have individual MNPs, which would have very different properties from the ones in assemblies. In any case, more work is required to elucidate this issue.

Apart the present experimental results, a few other experimental studies can be discussed in view of our findings. In Refs [24] and [25], Gudoshnikov *et al.* evidenced that increasing the aspect ratio of dense assemblies of MNPs increase their SAR, due to effect of the demagnetizing field. This is in agreement with our theoretical finding that anisotropic assemblies of MNPs displays an enhanced SAR compared to the isotropic ones. In Ref. [26], Alphandery *et al.* have shown that chains of MNPs are dominated by dipolar interactions, irrespective of the exact orientation of the MNP individual easy axis inside them, and that the squareness of their quasi-static hysteresis loop is enhanced by the presence of magnetic interactions. We show here that this finding is certainly true for very low anisotropy MNPs (the ones studied by the authors belong to this category) but cannot be generalized since large anisotropy MNPs maintain the characteristics of non-interacting MNPs. In Ref [27], Müller *et al.* observed very different SAR values between MNPs which were gelled with or without applying a magnetic field. Although no interpretation was provided by the authors, we attribute the increase of SAR in the samples textured by a large magnetic field to the formation of MNP chains or columns.

Some authors have attributed the increase of SAR values when particles are in a liquid compared to the ones found in a gel to Brownian motion [28]. This explanation should be proposed by the authors with caution because there is no specific contribution of brownian motion to magnetic hyperthermia, which could be separated from another contribution. SAR value is the result of the global MNP hysteresis loop. This point is clearly explained and illustrated in Ref. [23]. Our present study proposes an alternative plausible explanation to the increase of SAR in liquids, resulting from the formation of chains of MNPs. This hypothesis should be seriously envisaged by authors when such experimental result is found.

Finally, our theoretical results are consistent with theoretical works of Refs [3] and [9] since we similarly show that, in isotropic assemblies of MNPs, magnetic interactions decrease SAR compared to independent MNPs. We show an additional feature which is that, due to the formation of chains and columns in magnetic hyperthermia experiments, low anisotropy MNPs present enhanced SAR compared to independent NPs. This point was missed in previous theoretical works: in Refs [3], [8] and [10], only isotropic assemblies were studied; in Ref [9], in spite that anisotropic assemblies of low anisotropy MNPs near the superparamagnetic/ferromagnetic transition were studied, only a very limited increase of SAR was detected. This may be due to the fact that the effect of magnetic interactions in these assemblies also leads to an increase of coercive field (see Fig. 6) and that simulations in Ref [9] were performed at a small magnetic field which did not saturate the hysteresis loops of interacting NPs and masked the effect.

5. Conclusion

Our main finding can be summarized this way : the magnetic interactions in the chains of MNPs formed during hyperthermia experiments have a tendency to induce a uniaxial anisotropy, which increases the squareness of the hysteresis loop and thus MNP efficiency. This process is in competition with the MNP magnetocrystalline anisotropy, which has the tendency to decrease the squareness and efficiency of the MNPs toward the one of magnetically independent MNPs. This effect is then visible only for low-anisotropy MNPs saturated by the applied magnetic field. Since this configuration is precisely the one where MNPs are optimized for magnetic hyperthermia and display the highest SARs [1], this finding is important to interpret experiments. Chains of MNPs with a uniaxial anisotropy are the only way to reach the maximum possible SAR with a given magnetic material.

These results evidence the importance of considering the chains and columns of MNPs formed during the application of the magnetic field to interpret correctly magnetic hyperthermia. Future theoretical work on magnetic interactions should not miss this point. It would be very interesting to know to which extent such chains are formed during *in vivo* and *in vitro* experiments : in spite of their low overall concentration in tumors, MNPs are in some case grouped in intracellular compartments of cells and locally reach high concentration. These are conditions where magnetic interactions could be non-negligible and could then induce the formation of chains that lead to large SAR values.

**Acknowledgements :**
This work was supported by the Fondation InNaBioSanté and the Région Midi-Pyrenées.

**Appendix**

Determination of Equ. (2) follows the same principle as the determination of Equ. (1), which was previously described in [12]. In the ferromagnetic regime, the magnetic field dependence of SAR in hyperthermia measurements shows an abrupt increase, which occurs when the applied magnetic field exceeds the coercive field. Since the coercive field is not an intrinsic parameter of the MNPs but depends on the amplitude of the applied magnetic field, the determination of the exact value of this abrupt increase requires a dedicated study. For that purpose, SAR($\mu_0 H_{max}$) functions were calculated for different NP diameter in the case where their anisotropy axis is aligned with the magnetic field. The maximum slope of the SAR($\mu_0 H_{max}$) function occurs when $\mu_0 H_{max} = \mu_0 H_{CHyp}$. In Fig.9, $\dfrac{H_{CHyp}}{H_K}$ is plotted as a function of $\kappa$, where $\mu_0 H_K = \dfrac{2K_{ind}}{M_S}$ is the anisotropy field and $\kappa = \dfrac{k_B T}{K_{ind} V} \ln\left(\dfrac{k_B T}{4\mu_0 H_{max} M_S V f \tau_0}\right)$. It has been shown in previous studies that $\kappa$ is a good dimensionless parameter to describe the coercive field in the ferromagnetic regime [1, 12, 29]. Data in Fig. 9 were then fitted using variations of Equ. (2) with different pre-factors. The best fit occurs when the prefactor equals 1.85 (see Fig. 9), which permits to determine the final form of Equ. (2).

# Tables

| Samples | $d_0$ (nm) TEM | %$Fe_{2.2}C$ XRD | $M_S$ ($Am^2$/kg) SQUID | $\mu_0 H_{CHyp}$ (mT) hyperthermia | $\mu_0 H_C$ (mT) loops | $K_{eff}$ (kJ/$m^3$) hyperthermia | $K_{eff}$ (kJ/$m^3$) loops | $A$ (mJ/g) hyperthermia | $A$ (mJ/g) loops | $M_{sat}$ ($Am^2$/kg) loops | $M_R/M_{sat}$ loops | Squareness loops | Slope (a. u.) loops |
|---|---|---|---|---|---|---|---|---|---|---|---|---|---|
| Sample 1 FeCo | 12.8 | 0 | 140 | 8 | 8 | 45.7 1D | 45.7 1D | 1.05 | 1.1 | 33.0 | 0.92 | 0.69 | 544 |
| Sample 2 Fe(0) | 13.7 | 0 | 232 | 38 | 20 | 103 3D | 69 3D | 3.80 | 2.7 | 47.0 | 0.63 | 0.37 | 66 |
| Sample 3 $Fe_xC_y$ | 12.1 | 100 | 146 | > 60 | > 42 | >117 3D | >95 3D | 1.57 | 1.08 | 20.2 | 0.45 | 0.32 | 36 |
| Sample 4 Fe@$Fe_xC_y$ | 13.1 | 22 | 202 | 17 | 18 | 54.3 1D | 55.3 1D | 4.66 | 6.00 | 89.4 | 0.94 | 0.74 | 292 |
| Sample 5 Fe@$Fe_xC_y$ | 13.1 | 16 | 191 | 32 | 30 | 68.2 1D | 66.2 1D | 6.35 | 8.2 | 75.1 | 0.92 | 0.71 | 217 |
| Sample 6 Fe@$Fe_xC_y$ | 14.6 | 59 | 203 | 25 | 25 | 67.4. 3D | 67.4 3D | 7.9 | 8.6 | 97.8 | 0.83 | 0.56 | 116 |

**Table 1 : Summary of magnetic and hyperthermia properties for different samples:** Columns labeled with "hyperthermia" refers to data deduced from temperature measurements. Columns labeled with "loops" refers to data extracted from hysteresis loops measured at $f$ = 54 kHz and $\mu_0 H_{max}$ = 42 mT. "$d_0$": mean diameter determined by TEM. "%$Fe_{2.2}C$": phase fraction of $Fe_{2.2}C$ deduced from XRD. "$M_S$": saturation magnetization for $\mu_0 H_{max}$ = 5 T.deduced from SQUID measurements on powder and from microanalysis. "$\mu_0 H_{CHyp}$" coercive field deduced from the highest slope of the $SAR(\mu_0 H_{max})$ function. "$\mu_0 H_C$": coercive field deduced from hysteresis loops. "$K_{eff}$": effective anisotropy deduced from hyperthermia measurements or hysteresis loops. In the first (second) column, $\mu_0 H_{CHyp}$ ($\mu_0 H_C$) have been used to calculate $K_{eff}$ using the analytical equations presented in text ("1D" refers to axis aligned with the magnetic field and "3D" to randomly oriented axis). "$A$": losses per cycle deduced from hyperthermia and hysteresis loop measurements. "$M_{sat}$" : magnetization at saturation (see text). "$M_R/M_{sat}$" ratio between remnent and saturation magnetization. "Squareness" is calculated from Equ.(5). "Slope" is the slope of the hysteresis loop at the coercive field.

**Figures :**

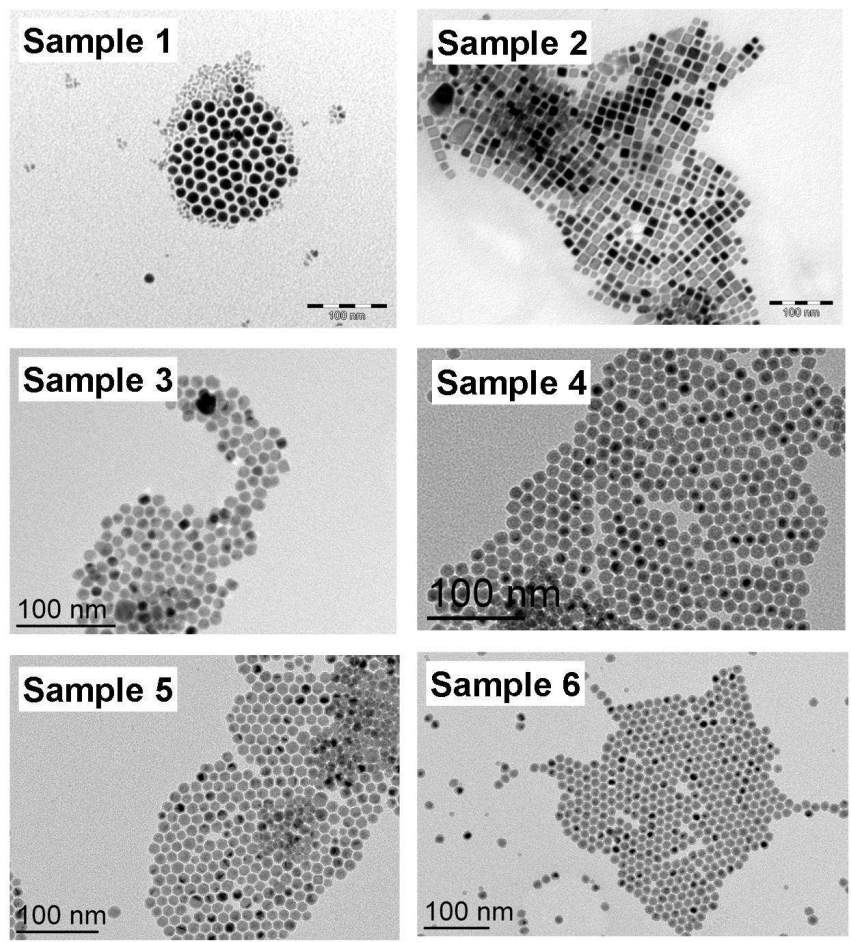

Figure 1: TEM micrographs of the samples. The length of the scale bar is 100 nm.

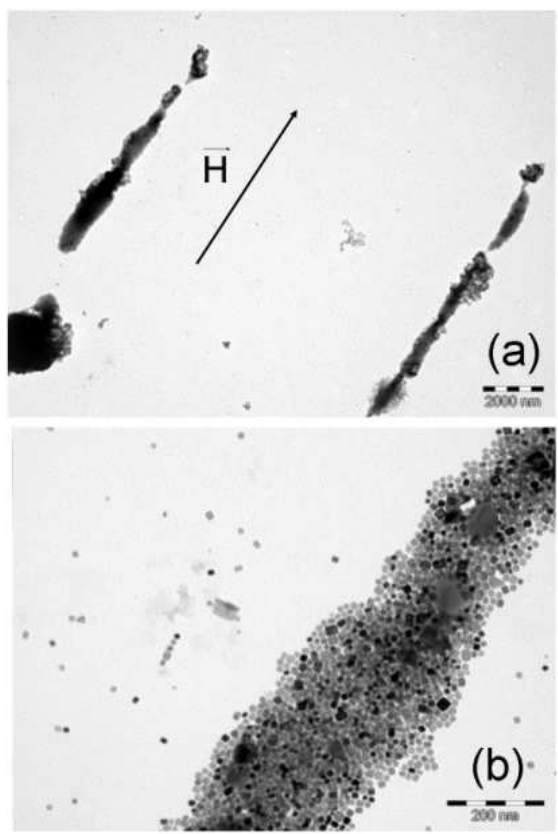

Figure 2: (a) and (b) TEM micrographs of experiments where a drop of colloidal solution containing MNPs of Fe was deposited on a grid of microscopy and let dry under the application of a magnetic field of 40 mT at 54 kHz. The length of the bar is (a) 2 µm (b) 200 nm.

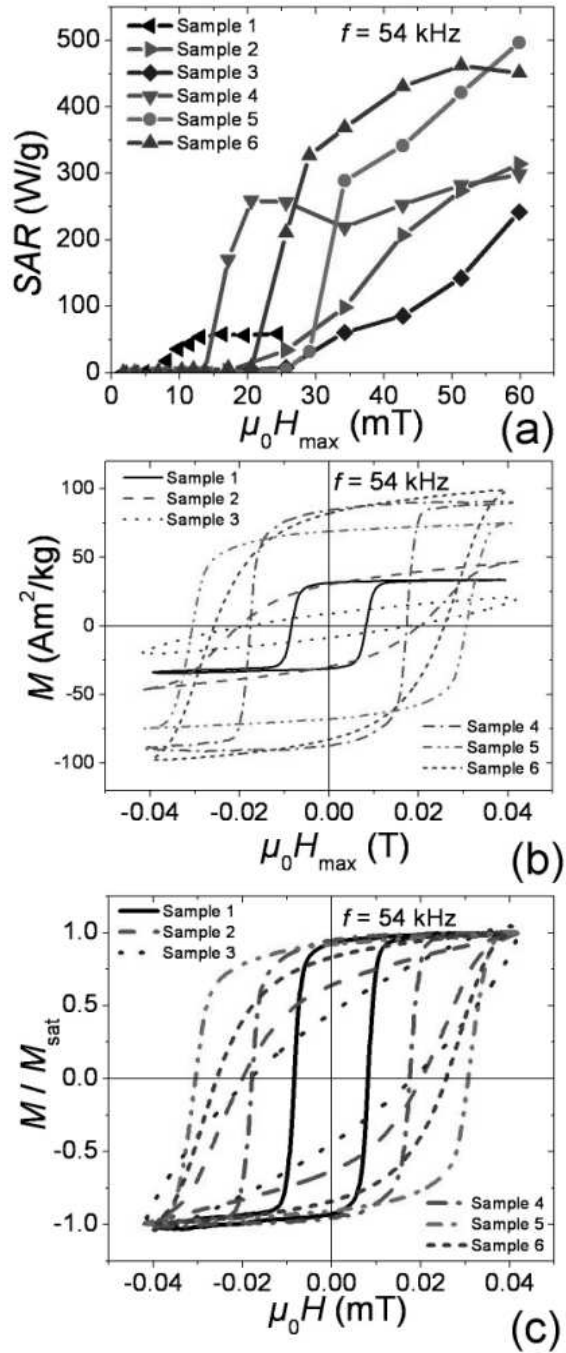

Figure 3 (color online): (a) SAR as a function of the magnetic field deduced from temperature measurements; $f = 54$ kHz (b) and (c) Hysteresis loop measurements performed at $f = 54$ kHz and $\mu_0 H_{max} = 42$ mT. In (c) the hysteresis loops are normalized by the magnetization value at 42 mT.

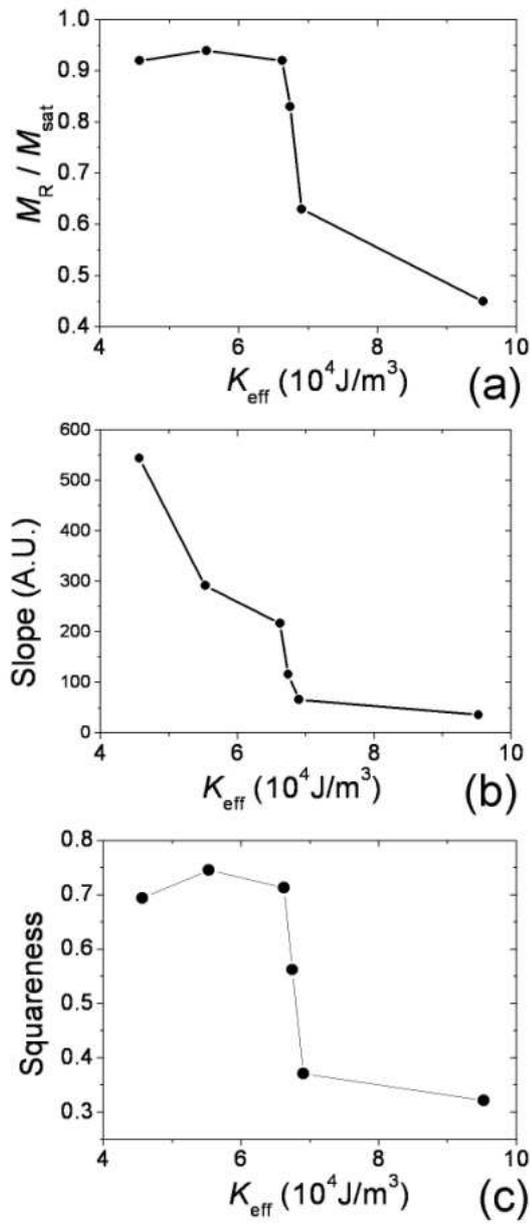

Figure 4 : Data extracted from the hysteresis loops shown in Fig. 3 and plotted as a function of the estimated $K_{eff}$ value of the different samples. The parameters plotted are (a) the normalized remanent magnetization $M_R/M_{sat}$, (b) the slope at the coercive field, (c) the squareness $S$.

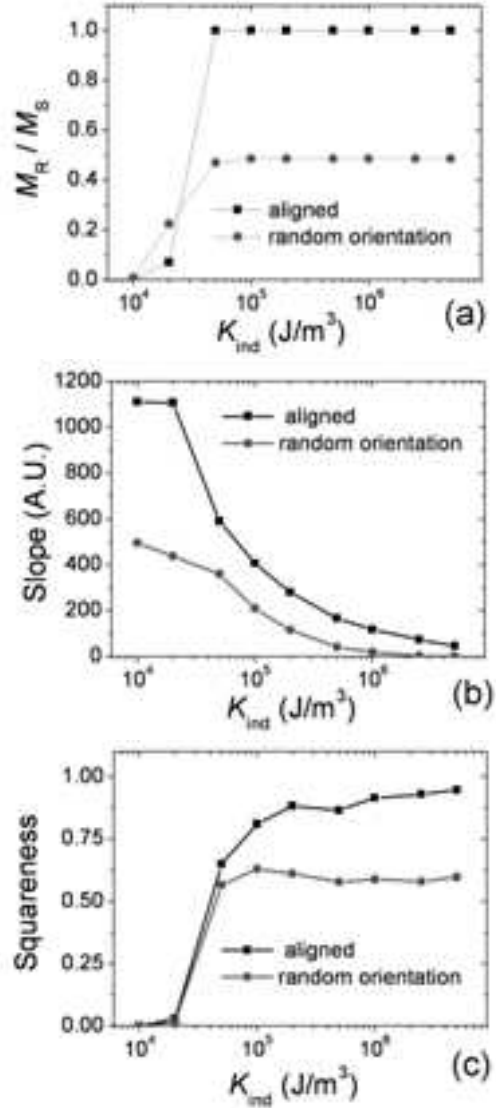

Figure 5 : Data extracted from numerical simulations of non-coupled single-domain MNPs. MNPs have either their anisotropy axis oriented with the magnetic fied or randomly oriented in space. Parameters are $\tau_0 = 5\times10^{-11}$ s, $M_S = 2\times10^6$ A m$^{-1}$, $T = 300$ K, $f = 54$ kHz and a varying $K_{ind}$. The graphs are (a) the normalized remanent magnetization $M_R/M_S$, (b) the slope at the coercive field and (c) the squareness. Examples of the corresponding hysteresis loops can be found in Ref [1].

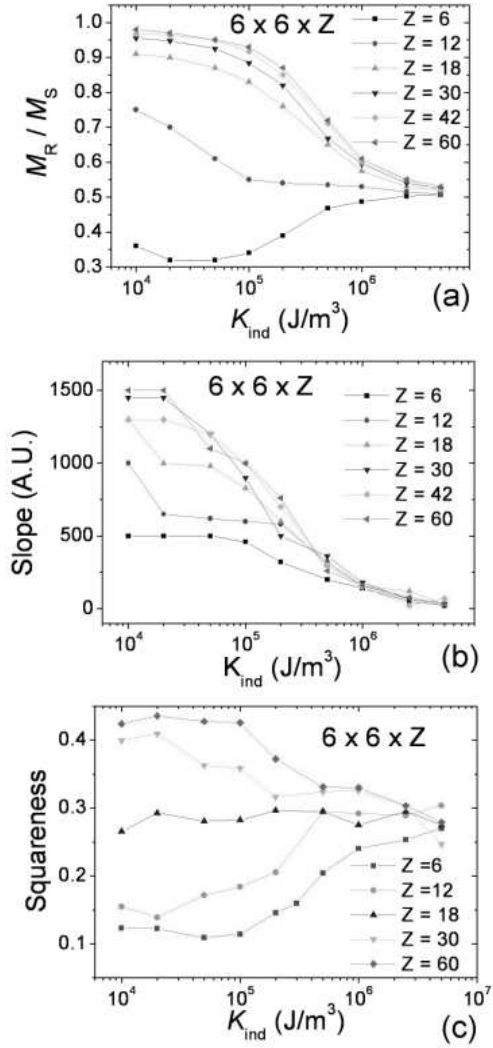

Figure 6: (color online) Data extracted from numerical simulations of magnetically coupled MNPs. 3D arrays of 6x6xZ MNPs have been simulated. Parameters are $M_S = 2 \times 10^6$ A m$^{-1}$, $D = 13$ nm, the center-to-center interparticle distance was 15 nm, $K_{ind}$ was varied. The data plotted are (a) the normalized remnant magnetization $M_R/M_S$, (b) the slope at the coercive field and (c) the squareness. Examples of the corresponding hysteresis loops are shown in Fig. 7.

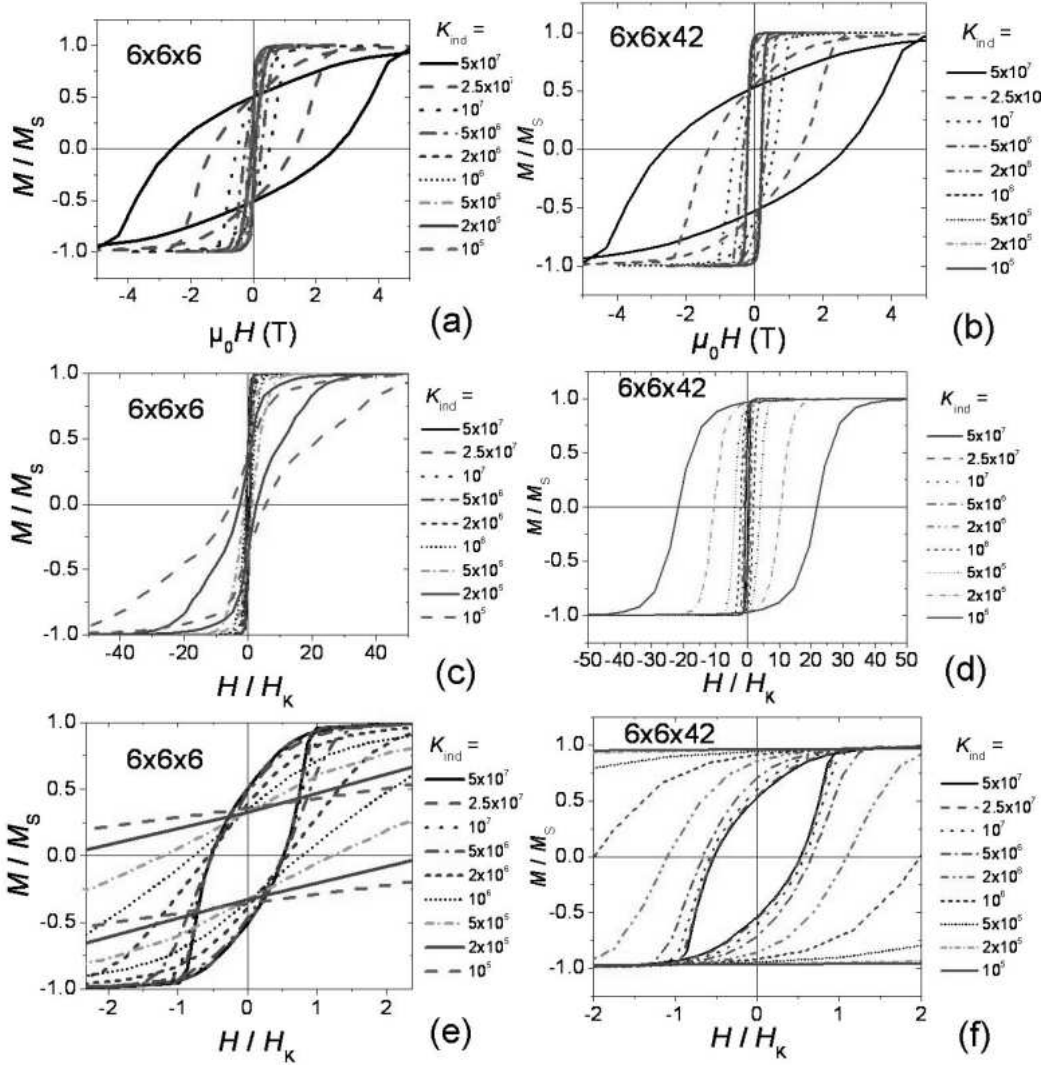

Figure 7: (color online) Numerical simulations of magnetically coupled MNPs. Simulation parameters are given in Fig. 6. The array size is (a) (b) (c) 6x6x6 and (d) (e) (f) 6x6x42. In (e) and (f) the magnetic field $\mu_0 H$ is normalized by the anisotropy field $\mu_0 H_K = \dfrac{2K_{ind}}{M_S}$.

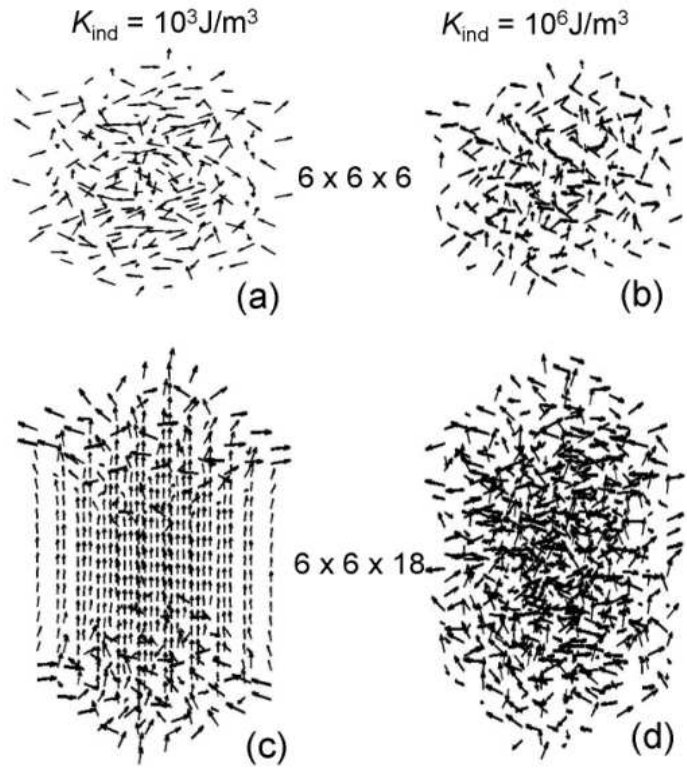

Figure 8: Magnetic configuration of the nanoparticles in the remanent state for two different values of aspect ratio and anisotropies. Each arrow indicates the magnetization direction of a MNP. (a) 6x6x6, $K_{ind} = 10^3 J/m^3$. (b) 6x6x6, $K_{ind} = 10^6 J/m^3$. (c) 6x6x18, $K_{ind} = 10^3 J/m^3$. (d) 6x6x18, $K_{ind} = 10^6 J/m^3$.

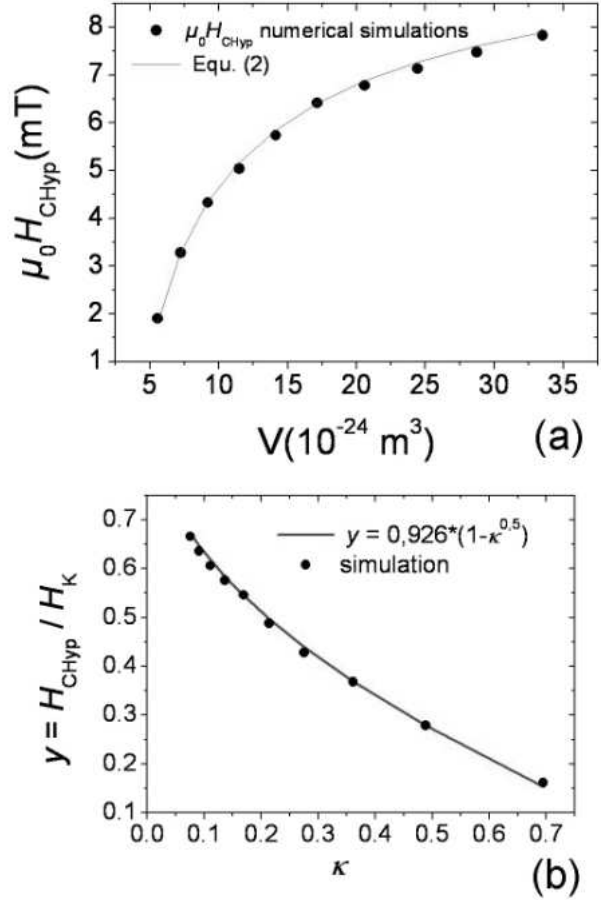

Figure 9: Normalized $\mu_0H_{CHyp}$ values plotted as a function of $\kappa$. Each $\mu_0H_{CHyp}$ value has been determined from the maximum of the derivative of SAR($\mu_0H_{max}$) functions. The numerical simulations were run with $K_{ind} = 1 \times 10^4$ J/m$^3$, $T = 300$ K, $f = 100$ kHz, $\tau_0 = 5 \times 10^{-11}$ s, $\mu_0H_{max}$ in the range 0-70 mT, and with a varying NP diameter ranging from 2 to 40 nm. Dots correspond to the values extracted from the numerical simulations and the solid line to the numerical solving of Equ. (2).